\documentclass[
    ,final            
  ]
  {aipproc}

\layoutstyle{8x11double}

\begin{document}

\title{Estimates for Very High Energy\\ Gamma Rays from Globular Cluster Pulsars}


\author{C. VENTER}{
  address={Unit for Space Physics, North-West University, Potchefstroom Campus, Private Bag X6001, Potchefstroom 2520, South Africa}, altaddress={The Centre for High Performance Computing, CSIR Campus, 15 Lower Hope Street, Rosebank, Cape Town, South Africa}
}

\author{O.C. DE JAGER}{
  address={Unit for Space Physics, North-West University, Potchefstroom Campus, Private Bag X6001, Potchefstroom 2520, South Africa}, altaddress={South African Department of Science and Technology, and National Research Foundation Research Chair: Astrophysics and Space Science}
}

\begin{abstract}
Low-Mass X-ray Binaries (LMXRBs), believed to be the progenitors of recycled millisecond pulsars (MSPs), occur abundantly in globular clusters (GCs). GCs are therefore expected to host large numbers of MSPs. This is also confirmed observationally. The MSPs continuously inject relativistic electrons into the ambient region beyond their light cylinders, and these relativistic particles produce unpulsed radiation via the synchrotron and inverse Compton (IC) processes. It is thus possible, in the context of General Relativistic (GR) frame-dragging MSP models, to predict unpulsed very high energy radiation expected from nearby GCs.
We use a period-derivative cleaned sample of MSPs in 47~Tucanae, where the effects of the cluster potential on the individual period derivatives have been removed. This MSP population is likely to have significant pair production inhibition, so that slot gaps and outer gaps are not expected to form in the pulsar magnetospheres. The utilisation of unscreened pulsar potentials is therefore justified, and fundamental tests for the predicted average single pulsar gamma-ray luminosities and associated particle acceleration are simplified. Using a Monte Carlo process to include effects of pulsar geometry, we obtain average injection spectra (with relatively small errors) of particles leaving the MSPs. These spectra are next used to predict cumulative synchrotron and IC spectra expected from 47~Tucanae, which is a lower limit, as no reacceleration is assumed.
We find that the IC radiation from 47~Tucanae may be visible for \textit{H.E.S.S.}, depending on the nebular field $B$ as well as the number of MSPs $N$ in the GC. Telescopes such as \textit{Chandra} and \textit{Hubble} may find it difficult to test the SR component prediction of diffuse radiation if there are many unresolved sources in the field of view. These results may be rescaled for other GCs where less information is available, assuming universal GC MSP characteristics.
\end{abstract}

\maketitle


\section{Introduction}
A total of 137 globular cluster (GC) pulsars have been discovered in 25~GCs\footnote{http://www.naic.edu/$\sim$pfreireGCpsr.html}, following the discovery of the first GC millisecond pulsar (MSP) in M28 \citep{Lyne87}. Low-Mass X-ray Binaries (LMXRBs), believed to be the progenitors of recycled MSPs \citep{Alpar82}, occur abundantly in GCs. GCs are therefore expected to host large numbers of MSPs, up to $\sim200$ MSPs or more \citep{Ivanova05,BS07}. Indeed, GC MSP spin properties seem consistent with the recycling scenario \citep{Ransom08b}.

Terzan~5, 47~Tucanae, and M28 collectively contain nearly half of all GC pulsars, housing 33, 23, and 11 pulsars respectively \citep{Ransom08b}. GC MSPs are sources of relativistic electrons, which are continuously being injected into the ambient region beyond the MSPs' light cylinders. These relativistic particles produce high-energy emission via synchrotron radiation (SR) and inverse Compton scattering (ICS) on bright starlight photons as well as on the cosmic microwave background (CMB). 

In this paper, we calculate the cumulative injection spectrum and resulting unpulsed SR and ICS fluxes, using a population of 13~MSPs in 47~Tucanae, with corrected values of their period-derivatives \citep{Bogdanov06b}. GC gamma-ray visibility is also discussed. We use a more refined injection spectrum, originating from a General Relativistic (GR) frame-dragging MSP model (e.g.\ \citep{Muslimov92,MH97,HM98}), than that assumed by \citep{BS07}, and our calculations are complementary to pulsed gamma-ray flux predicted by \citep{Venter_ApJL08}. As we only consider particles originating from MSP magnetospheres, with no further acceleration, our calculations should be viewed as lower limits to the expected TeV flux.

\begin{figure}
  \includegraphics[height=.28 \textheight]{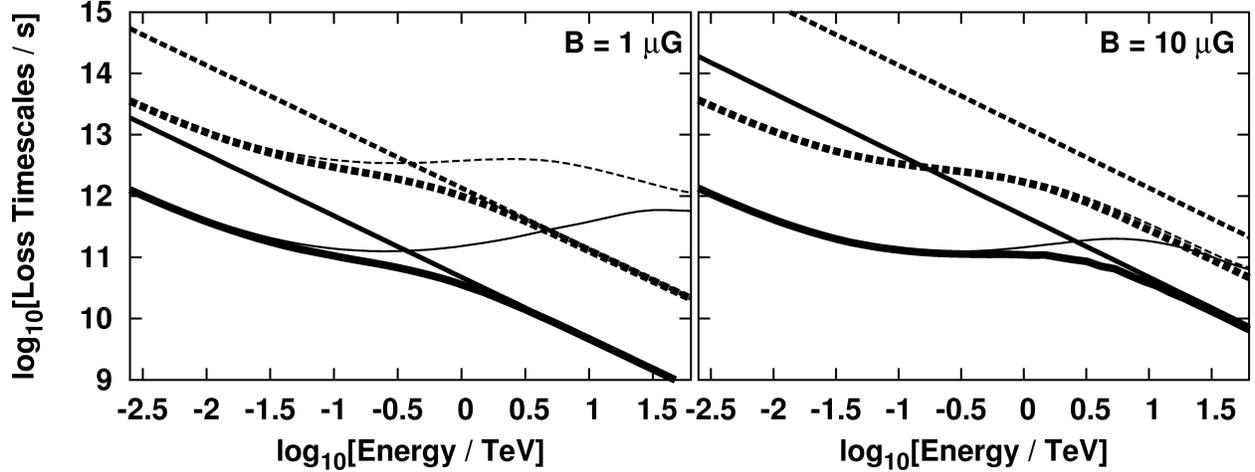}
  \caption{Loss timescales (left panel for $B=1\,\mu$G, right panel for $B=10\,\mu$G). Solid lines represent $Z=0$, dashed lines $Z=1$, thin lines $\tau_{\rm rad}$, intermediate straight lines $\tau_{\rm esc}$, and thick lines $\tau_{\rm eff}$ (see text for details).\label{fig:tau}}
\end{figure}

\section{Injection Spectrum Calculation}
\label{sec:Q}
As in \citep{Venter_ApJL08,Venter_PhD08}, we use a population of 13~MSPs in 47~Tucanae, with corrected values of their period-derivatives $\dot{P}$ \citep{Bogdanov06b}. We calculate the injection spectrum $Q^i$ of electrons leaving each MSP $i$, with $i=1,\cdots,13$, by binning the number of primary electrons leaving a stellar surface patch and moving along a B-line according to $E_{\rm e}^{\rm LC}\equiv\gamma_{\rm LC}m_ec^2$, the residual electron energy at the light cylinder, divided by energy bin size. 

\begin{figure}
  \includegraphics[height=.28\textheight]{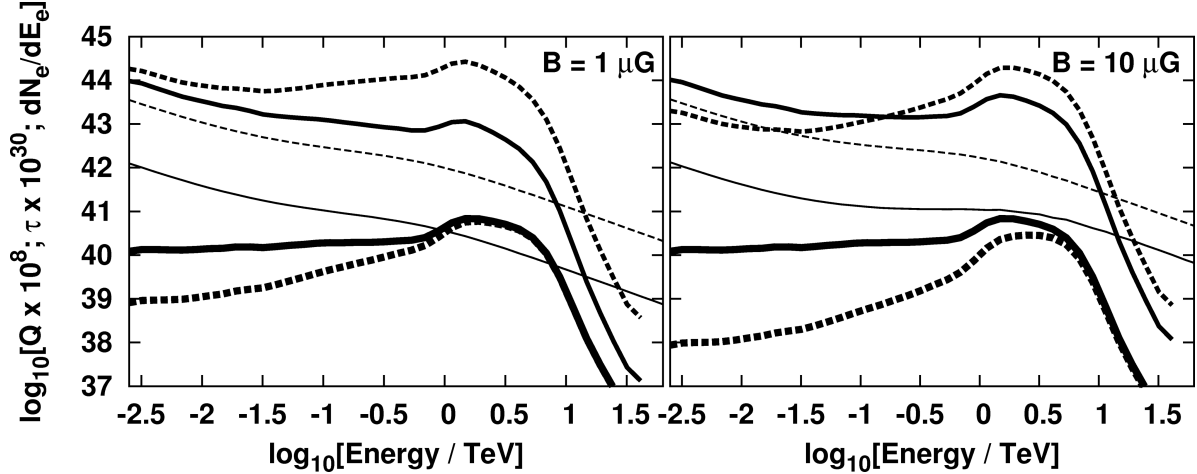}
  \caption{Bottom thick lines: injection spectra $Q\times10^8$ (/TeV/s); thin lines: $\tau_{\rm eff}\times10^{30}$ (s); top intermediate lines: steady-state electron spectra $dN_e/dE_e$ (/TeV) (left panel for $B=1\,\mu$G, right panel for $B=10\,\mu$G). Solid lines represent $Z=0$, dashed lines $Z=1$. We used $dN_e/dE_e=<Q^j>_{\rm cum}\times\tau_{\rm eff}$.\label{fig:dNdE}}
\end{figure}

We next randomly choose $N=100$ MSPs (with random inclination angles $\chi$), and sum their particle spectra to obtain a million cumulative spectra from the Monte Carlo process:
\begin{equation}
Q^j_{\rm cum} = \sum_{k=1}^{N=100}\frac{d\dot{N}^k_{\rm e}}{dE_{\rm e}}(E_{\rm e},\chi),
\end{equation}
where for each index $j=1,2,...,N_t=10^6$, a total of $N=100$ particle spectra (randomly sampled from 104 spectra: 13 MSPs $\times$ 8 values of $\chi$, $\chi=10^\circ,\,20^\circ,\cdots80^\circ$) are summed. We therefore oversample from~13 to 100~pulsar particle spectra, to obtain the cumulative injection spectrum for $N=100$~MSPs into the GC 47~Tucanae. Note that the single MSP particle spectra $Q^i$ are not functions of observer angle $\zeta$, so that the relative errors of the average cumulative particle spectrum $<Q^j_{\rm cum}>$ are smaller than for the average cumulative gamma-ray spectrum $<(dN_\gamma/dE)^j_{\rm cum}>$ at earth, which was calculated in \citep{Venter_ApJL08}. The relative uncertainty of $<Q^j_{\rm cum}>$ is due to the different values of $P$ and $\dot{P}$, as well as different inclination angles $\chi$, of the MSP population members. The thick solid lines in Figure~\ref{fig:dNdE} indicate $<Q^j_{\rm cum}>$.

\section{Unpulsed Gamma-Ray Flux}
\label{sec:Flux}
We divide the region where unpulsed radiation is generated into two zones: `zone 0' ($Z=0$) reaching from $r=0$ to $r=r_{\rm c}$, with $r_{\rm c}$ the core radius, and `zone 1' ($Z=1$), reaching from $r=r_{\rm c}$ to $r=r_{\rm hm}$, with $r_{\rm hm}$ the half mass radius. Using the cluster core formula (Eq.~5 in \citep{BS07}), we find an energy density $u_{\rm rad}\sim3000$ eV/cm$^3$ for $Z=0$, and $u_{\rm rad}\approx L_{\rm GC}/(4\pi \overline{r}^2 c)\sim100$ eV/cm$^3$ for $Z=1$ for the bright starlight component (corresponding to temperature $T\approx4500$~K), with $L_{\rm GC}$ the observed cluster luminosity, and $\overline{r}=(r_{\rm c}+r_{\rm hm})/2$. We use similar parameters as \citep{BS07}: distance $d=4.5$~kpc, $r_{\rm c}=0.58$~pc, $r_{\rm hm}=3.65$~pc, and $L_{\rm GC}=7.5\times10^5L_\odot$. For the CMB component, we use $u_{\rm CMB}\sim0.27$~eV/cm$^3$ (corresponding to $T\approx2.76$~K). 

The radiation loss, escape, and effective timescales (Figure~\ref{fig:tau}) are given by
\begin{eqnarray}
\tau_{\rm ICS} & = & \frac{E_{\rm e}}{\dot{E}_{\rm e,\,ICS}}\\
\tau_{\rm SR} & = & \frac{E_{\rm e}}{\dot{E}_{\rm e,\,SR}}\propto B^{-2}E_{\rm e}^{-1}\\
\tau_{\rm rad} & = & \frac{E_{\rm e}}{\dot{E}_{\rm e,\,SR}+\dot{E}_{\rm e,\,ICS}}\\
\tau_{\rm esc} & = & \frac{r_{\rm esc}^2}{2\kappa_{\rm Bohm}}\propto BE_{\rm e}^{-1} r_{\rm esc}^2,
\end{eqnarray}
with $\kappa_{\rm Bohm}=cE_{\rm e}/(3qB)$ and $r_{\rm esc}=r_{\rm c}$ for $Z=0$, $r_{\rm esc}=r_{\rm hm} - r_{\rm c}$ for $Z=1$, and \citep{Zhang08}
\begin{equation}
\tau^{-1}_{\rm eff}\approx\tau^{-1}_{\rm esc}+\tau^{-1}_{\rm rad}. 
\end{equation}
The left panel of Figure~\ref{fig:tau} is for a nebular field $B=1\,\mu$G, while the right panel is for $B=10\,\mu$G. Solid lines represent $Z=0$, and dashed lines $Z=1$. 

For $B=$~$10\mu$G, $\tau_{\rm esc}\propto B$ is $\sim10$ times larger than for $B=1$~$\mu$G. The bright starlight component dominates $\tau_{\rm rad}$ at low energies for all B-field strengths, while the CMB component dominates for $B=1$~$\mu$G at high energies, and $\tau_{\rm rad}$ is $\sim10$ times lower for the case of $B=10$~$\mu$G than for the case of $B=1$~$\mu$G at high energies. When $B=1$~$\mu$G and $B=10$~$\mu$G, $\tau_{\rm rad}\ll\tau_{\rm esc}$ at small $E_{\rm e}$, while $\tau_{\rm rad}\gg\tau_{\rm esc}$ at large $E_{\rm e}$ for $B=1\mu$G. Thus $\tau_{\rm eff}\approx\tau_{\rm rad}$ at small $E_{\rm e}$ for all field strengths, while $\tau_{\rm eff}\approx\tau_{\rm esc}$ at large $E_{\rm e}$ for $B=1$~$\mu$G, but $\tau_{\rm eff}\approx\tau_{\rm rad}$ for $B=10$~$\mu$G and $Z=1$. 

The steady-state electron spectrum is calculated using 
\begin{equation}
\frac{dN_{\rm e}}{dE_{\rm e}}=<Q^j_{\rm cum}>\tau_{\rm eff} 
\end{equation}
for each zone. Figure~\ref{fig:dNdE} indicates the (scaled) injection spectrum for $Z=0$ by thick solid lines (average of Eq.~1). We divide $dN/dE_{\rm e}$ of $Z=0$ by $\tau_{\rm esc}$ of $Z=0$, and obtain the thick dashed lines for the injection spectrum of $Z=1$.The product of $\tau_{\rm eff}$ (thin lines indicating scaled $\tau_{\rm eff}$) with the injection spectra (thick lines) give the steade-state electron spectra for each zone (lines of intermediate thickness). As previously, the left panel of Figure~\ref{fig:dNdE} is for $B=1\,\mu$G, while the right panel is for $B=10\,\mu$G. Solid lines represent $Z=0$, and dashed lines $Z=1$. 

\section{Discussion and Conclusions}
\label{sec:concl}
\begin{figure}
\includegraphics[height=.3\textheight]{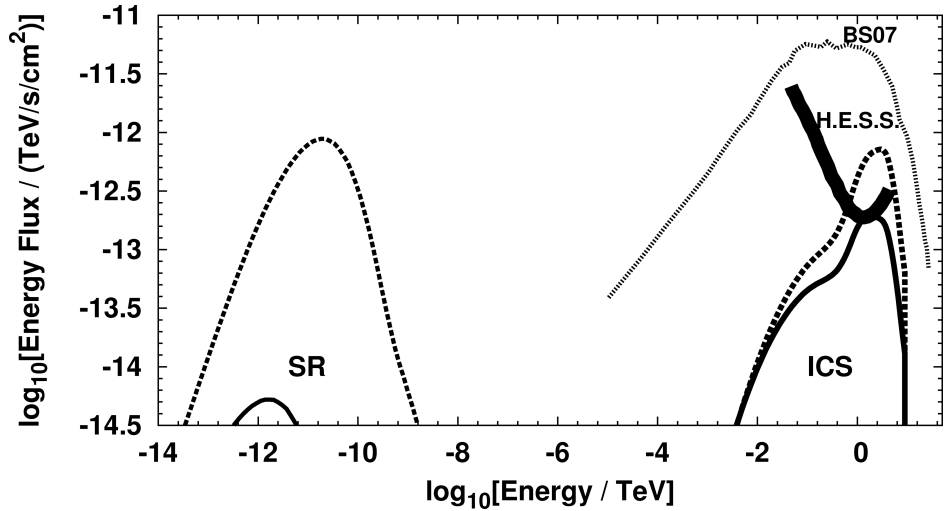}
\caption{Unpulsed SR (thin lines) and ICS (thick lines) components (solid lines: $B=1$~$\mu$G, dashed lines: $B=10$~$\mu$G; summation of $Z=0$ and $Z=1$ spectra), with the latter consisting of a bright starlight and CMB component. Also shown are a prediction from \citep{BS07} (short-dashed line), and the \textit{H.E.S.S.} sensitivity \citep{Hinton04}.\label{fig:Diff}}
\end{figure}
The predicted SR and ICS fluxes are larger for $B=10\mu$G than for $B=1\mu$G (the SR fluxes differ by about 2 orders of magnitude). The SR component occurs in the optical / X-ray waveband, and the ICS (scattering starlight and CMB), in the gamma-ray waveband. 

Our predictions for the IC flux are notably smaller than those of \citep{BS07}. This is mainly attributed to the fact that \citep{BS07} assumed an average spindown luminosity per MSP of $\sim10^{35}$ erg/s, while we find $\sim10^{34}$ erg/s when using the population of 13 MSPs with known `cleaned' $\dot{P}$ (although we both assumed $N=100$ GC members). Also, \citep{BS07} used an average particle conversion efficiency of $\eta=0.01$, which is quite close to our numerical result of $\eta=0.0074$ obtained in \citep{Venter_ApJL08}. Lastly, our spectral shapes differ, as \citep{BS07} approximated the injection spectra by power laws with different spectral indices and cut-offs, while we calculated the cumulative injection spectrum using the GR frame-dragging model as well as an actual GC population of MSPs. 

Due to the very bright starlight component in the GC core, the ICS from 47~Tucanae may be visible for \textit{H.E.S.S.}, depending on the nebular field $B$ as well as the number of MSPs $N$ in the GC. (A lowest limit may be obtained by multiplying the graphs by 0.23, thereby assuming $N=23$). Telescopes such as \textit{Chandra} and \textit{Hubble} may find it difficult to test the SR component prediction of diffuse radiation if there are many unresolved sources in the field of view. 

The results above 1~TeV may be roughly scaled for other GCs, where less information is available, by the factor
\begin{equation}
x\approx\left(\frac{N}{100}\right)\left(\frac{d_{\rm Tuc}}{d}\right)^2\left(\frac{\left<u\right>}{\left<u_{\rm Tuc}\right>}\right),
\end{equation}
with $<u>\sim L_{\rm GC}/(4\pi r_{\rm hm}^2)$, assuming universal GC MSP characteristics. As only MSPs were considered as sources of relativistic particles, and no further particle acceleration has been assumed, our estimates for the TeV flux should be viewed as lower limits.


\begin{theacknowledgments}
This publication is based on research supported by the South African National Research Foundation and the SA Centre for High Performance Computing.
\end{theacknowledgments}

\bibliographystyle{aipproc}   
\bibliography{Heidelberg}
\end{document}